\documentstyle[prl,aps,12pt]{revtex}
\begin{document}
\draft
\title{Mass Renormalization for Time Dependent Correlation Functions in Shibata-Hashitsume's Projection Operator Method}
\author{T.Koide,M.Maruyama and F.Takagi}
\address{Department of Physics,Tohoku University}

\maketitle

\begin{abstract}
 We study the time development of correlation functions at both zero and finite temperature 
with Shibata-Hashitsume's projection operator method
and carry out the renormalization of ultraviolet divergence 
that appears in a time-dependent frequency shift, using a mass counter term.
A harmless divergence of log $t$-type 
remains at an initial time $t=0$ after the lowest order renormalization.
\end{abstract}


Calculating various physical quantities 
in a non-equilibrium system is a very interesting problem. 
However, it is well-known that there are various problems 
in calculating time dependent physical quantities, e.g., 
``the surface divergences''\cite{ref:14} and the short time behavior 
of an unstable system which is related to 
``Quantum Zeno's paradox''\cite{ref:15}.
In these problems, one encounters peculiar divergences 
which do not appear in usual S-matrix theory where the relevant 
time interval is [$-\infty$,$\infty$]. 
To avoid these divergences, for example, diffuse time boundaries 
have been introduced\cite{ref:14}\cite{ref:15}.

In this paper, we show first that a divergence appears 
at the initial time $t=0$ 
in correlation functions obtained by using Shibata-Hashitsume's 
projection operator method 
\cite{ref:10}\nocite{ref:11}--\cite{ref:12} and then show that 
it becomes harmless in calculating some physical quantities 
in a super-renormalizable model 
by mass renormalization.
Shibata-Hashitsume's 
projection operator method is one of 
various formulations of non-equilibrium statistical mechanics: 
e.g. linear response theory \cite{ref:1}, 
Caldeira-Leggett's theory \cite{ref:2},
closed time path method \cite{ref:3}\cite{ref:4},
projection operator method\cite{ref:7}\nocite{ref:8}\nocite{ref:9}\nocite{ref:10}\nocite{ref:11}--\cite{ref:12}, 
and so on \cite{ref:5}\cite{ref:6}.

Advantages of Shibata-Hashitsume's projection operator method 
in our analysis 
are that its generic formulation is independent of the explicit form of the projection operator 
and a non-perturbative effect can be taken into account by solving 
the Heisenberg equation of motion obtained by a projection.

Our strategy is to study quantum field theory 
as an initial value problem, 
that is, we prepare an initial state and determine the time evolution by 
solving the Heisenberg equation
\begin{eqnarray}
  \frac{d}{dt}O(t) &=& i [ H , O(t) ] \label{eqn:hei}\\
                   &=& iLO(t),  
\end{eqnarray}
where $L$ is the Liouville operator.
The Heisenberg equation has complete information on the time development 
of the operator,
but is difficult to solve exactly.
So it is necessary to make some approximations.
For this purpose, we introduce generic projection operators $P$ and $Q$ 
which have the following general properties: 
\begin{eqnarray}
      P^2 &=& P, \\
      Q &=& 1 - P, \\
      PQ &=& QP = 0. 
\end{eqnarray}
By these projection operators, we can derive the following two equations.
\begin{eqnarray}
  \frac{d}{dt}e^{iLt}P &=& e^{iLt}PiLP + e^{iLt}QiLP, \label{eqn:int1}\\
  \frac{d}{dt}e^{iLt}Q &=& e^{iLt}PiLQ + e^{iLt}QiLQ. \label{eqn:int2}
\end{eqnarray}
We can solve Eq.(\ref{eqn:int2}) for $e^{iLt}Q$ 
and substitute the solution into Eq.(\ref{eqn:int1}).
Then we can transform the Heisenberg equation (\ref{eqn:hei}) 
into the following form:
\begin{eqnarray}
     \frac{d}{dt}O(t) &=& (e^{iLt}P + Qe^{iLQt})\frac{1}{1-\Sigma(t)}iLO(0) 
\label{eqn:genmitu},
\end{eqnarray}
where
\begin{eqnarray}
     \Sigma(t) &\equiv& \int^{t}_{0}d{\tau}e^{-iL(t-\tau)}PiLQe^{iLQ(t-\tau)}. 
\end{eqnarray}

Now the total Hamiltonian 
can be divided into three parts: 
system(s) which we are interested in, 
environment(e) and interaction(se) between system and environment:
\begin{eqnarray}
      H = H_{s} + H_{e} + H_{se}.
\end{eqnarray}
Let us specify the projection operator.
We are interested in the detailed behavior of the system, but not in that 
of the environment.
Therefore, we assume that 
\begin{eqnarray}
        \rho = \rho_{s}\rho_{e},
\end{eqnarray}
where $\rho$,$\rho_{s}$ and $\rho_{e}$ 
are the density matrices of the total system, 
the system which we are interested in 
and the environment, all given at $t = 0$, respectively, 
and then define the projection operator as 
\begin{eqnarray}
      P O &=& {\rm Tr}[\rho_{e}O] \equiv \langle O \rangle_{e}
\end{eqnarray}
for any operator $O$.
By this projection, we can replace an operator of the environment with a 
c-number.
Furthermore, we consider, for simplicity, a special case where 
$\rho_{e}$ has only diagonal components when it is expanded in terms of
the eigenstates of $H_{e}$.
Using these properties of the projection operator, 
we have the following relations which play an important role in the 
following calculations:
\begin{eqnarray}
   PL_{s} &=& L_{s}P,\\
   PL_{e} &=& L_{e}P = 0,
\end{eqnarray}
where $L_{a} O = [H_{a},O]$ for a = s,e,se.
Then, $\Sigma(t)$ can be expressed as 
\begin{eqnarray}
      \Sigma(t) &=& (1 - R(t)W(t))Q, 
\end{eqnarray}
where 
\begin{eqnarray}
      R(t) &=& 1 + \sum_{n=1}^{\infty}(-i)^n \int^{t}_{0}dt_{1}
\int^{t_{1}}_{0}dt_{2} \cdots \int^{t_{n-1}}_{0}dt_{n} \tilde{L}_{se}(t_{n})\tilde{L}_{se}(t_{n-1}) \cdots \tilde{L}_{se}(t_{1}) \\
      W(t) &=& 1 + \sum_{n=1}^{\infty} i^n \int^{t}_{0}dt_{1}
\int^{t_{1}}_{0}dt_{2} \cdots \int^{t_{n-1}}_{0}dt_{n} Q\tilde{L}_{se}(t_{1})Q\tilde{L}_{se}(t_{2}) \cdots Q\tilde{L}_{se}(t_{n})Q.
\end{eqnarray}
and
\begin{eqnarray}
\tilde{L}_{se}(s) &\equiv& e^{-i(L_{s}+L_{e})s}L_{se}e^{i(L_{s}+L_{e})s}. \nonumber 
\end{eqnarray}
These formula provide a systematic perturbative expansion.

Finally we carry out a perturbative expansion of $(1-\Sigma(t))^{-1}$ 
up to the first order of the interaction $H_{se}$.
We get the equation
\begin{eqnarray}
      \frac{d}{dt}O(t) &=& e^{iLt}PiLO(0) \nonumber \\
      &-& e^{iLt}\int^{t}_{0}
 dse^{-iL_{s}s}Pe^{-iL_{e}s}L_{se}e^{i(L_{s}+L_{e})s}QLO(0) \nonumber \\
            &+& Qe^{iLQt}iLO(0) \label{eqn:spe}.
\end{eqnarray}
Notice that this is a perturbative expansion in a differential equation, 
so this is not a simple perturbation.
The solution contains a certain kind of higher order contributions 
systematically in the sense of ordinary perturbation theory.

We apply Eq.(\ref{eqn:spe}) to $\sigma$-$2\pi$ model defined by 
the Hamiltonian density 
\begin{eqnarray}
h(x)&=& \frac{1}{2}\{ \Phi(x)\Phi(x) 
+ \nabla\phi(x)\nabla\phi(x) + m_{s}^2(t_{0})\phi^2(x) \} \label{eqn:sig} \nonumber \\
  &+& \frac{1}{2}\{ \vec{\Pi}(x)\vec{\Pi}(x) 
+ \nabla\vec{\pi}(x)\nabla\vec{\pi}(x) 
+ m_{\pi}^2\vec{\pi}^2(x) \} \label{eqn:pi}\nonumber \\
  &+& g\phi(x)\vec{\pi}(x)\vec{\pi}(x) 
-\frac{1}{2}{\delta}m_{s0}^2(t_{0})\phi^2(x), \label{eqn:int}
\end{eqnarray}
where
\begin{eqnarray}
  &&m_{s}^2(t_{0}) =  m_{s0}^2 + {\delta}m_{s0}^2(t_{0}),
\end{eqnarray}
and $m_{s0}$ is the bare mass of $\sigma$ meson, $\delta m_{s0}^2(t_{0})$ is the counter term, $t_{0}$ is the time at which we renormalize the divergence, and $\Phi(x)$, $\Pi(x)$ are the conjugate fields of $\phi(x)$, $\pi(x)$ respectively.
Now we want to evaluate the time dependence of $\sigma$ meson field by regarding the $\pi$ meson field as environment.
Thus the first, the second and the third lines 
on the r.h.s. of (\ref{eqn:int}) are identified 
with $H_{s}$, $H_{e}$ and $H_{se}$, respectively.
The $\sigma$ meson field is expanded as 
\begin{eqnarray}
  \phi(x) = \sum_{\vec{k}}\frac{1}{\sqrt{ 2V\omega_{s}(\vec{k}) }}\{ c_{\vec{k}}(t)e^{i\vec{k}\vec{x}} +  c^{\dagger}_{\vec{k}}(t)e^{-i\vec{k}\vec{x}}\}, \\
  \Phi(x) = -i\sum_{\vec{k}}\sqrt{ \frac{\omega_{s}(\vec{k})}{2V} }\{ c_{\vec{k}}(t)e^{i\vec{k}\vec{x}} -  c^{\dagger}_{\vec{k}}(t)e^{-i\vec{k}\vec{x}}\}, 
\end{eqnarray}
where
\begin{eqnarray}
        \omega_{s}(\vec{k}) &=& \sqrt{\vec{k}^2 + m_{s}^2(t_{0})},
\end{eqnarray}
and $V$ is the volume of the total system. 
We take the limit $V \longrightarrow \infty$ 
at the end of the calculation.

Equation of motion for the creation operator 
of the $\sigma$ meson is then given by 
\begin{eqnarray}
\frac{d}{dt} \langle c_{\vec{k}}^{\dagger}(t) \rangle &=& i\omega_{s}(\vec{k})\langle c_{\vec{k}}^{\dagger}(t) \rangle + i\rho(\vec{k}) \nonumber \\
    &+& \frac{{\rm Re}~A(\vec{k},t)}{\omega_{s}(\vec{k})}( \langle c^{\dagger}_{\vec{k}}(t) \rangle - \langle c_{-\vec{k}}(t) \rangle ) \nonumber \\   
    &+& \frac{i}{\omega_{s}(\vec{k})}( {\rm Im}A(\vec{k},t) - \frac{1}{2}{\delta}m_{s0}^2(t_{0}) ) ( \langle c^{\dagger}_{\vec{k}}(t) \rangle + \langle c_{-\vec{k}}(t) \rangle ),\label{eqn:sigma2pi} 
\end{eqnarray}
where
\begin{eqnarray}
        \rho(\vec{k}) &=&  \frac{g}{\sqrt{2V\omega_{s}(\vec{k})}}\int_{V}d^3\vec{x} \langle \vec{\pi}(\vec{x},0)\vec{\pi}(\vec{x},0) \rangle_{e} e^{i\vec{k}\vec{x}},   \\
      A(\vec{k},t) &=&  -\frac{g^2}{2V} \int^{t}_{0}ds\int_{V}{d^3}\vec{x}'{d^3}\vec{x}
      \langle [\vec{\pi}(\vec{x}',0)\vec{\pi}(\vec{x}',0),\vec{\pi}(\vec{x},s)\vec{\pi}(\vec{x},s) ] \rangle_{e} \label{eqn:A}\nonumber\\
      && \times e^{i\vec{k}\vec{x}}e^{-i\vec{k}\vec{x'}}e^{-i\omega_{s}(\vec{k})s},\\
       \vec{\pi}(\vec{x},t) &=& e^{iL_{e}t}\vec{\pi}(\vec{x}),
\end{eqnarray}
and $\langle~~\rangle_{e}$ and $\langle~~\rangle$ mean to take expectation values by using Tr $\rho_{e}$ and Tr $\rho$, respectively.
Here we consider the case where, 
at the initial time $t = 0$, $\pi$ meson field is 
in a thermal free gas phase of temperature $T$ 
while $\sigma$ meson field is in a coherent state.
Therefore, the expectation values of $\pi$ mesons 
and $\sigma$ mesons are taken 
for the thermal and the coherent states, respectively. 
The Fourier transform $\rho(\vec{k})$ of $\pi$-$\pi$ correlation function 
is the mean field felt by $\sigma$ mesons and is of the 1st order in $g$, 
while the $2\pi$-$2\pi$ correlation function $A(\vec{k},t)$ 
is of the second order in $g$.
The real and the imaginary parts of $A(\vec{k},t)$ are related to 
a damping rate 
and a frequency shift of $\sigma$ meson, respectively.
As seen in (\ref{eqn:sigma2pi}), 
equation for $\langle c_{\vec{k}}^{\dagger}(t) \rangle$ 
couples with $\langle c_{-\vec{k}}(t) \rangle$. 
Therefore, we need the equation for 
$\langle c_{-\vec{k}}(t) \rangle$:
\begin{eqnarray}
\frac{d}{dt} \langle c_{-\vec{k}}(t) \rangle 
   &=& -i\omega_{s}(\vec{k})\langle c_{-\vec{k}}(t) \rangle - i\rho(\vec{k}) \nonumber \\
   &-& \frac{{\rm Re}~A(\vec{k},t)}{\omega_{s}(\vec{k})}( \langle c^{\dagger}_{\vec{k}}(t) \rangle - \langle c_{-\vec{k}}(t) \rangle )\nonumber \\ 
    &-& \frac{i}{\omega_{s}(\vec{k})}({\rm Im}A(\vec{k},t) - \frac{1}{2}{\delta}m_{s0}^2(t_{0}))( \langle c^{\dagger}_{\vec{k}}(t) \rangle + \langle c_{-\vec{k}}(t) \rangle ).\label{eqn:sigma2pi2}
\end{eqnarray}
From Eq.(\ref{eqn:A}), we obtain 
\begin{eqnarray}
A(\vec{k},t) &=&\frac{3ig^2}{16{\pi}^{2}|\vec{k}|}\int^{t}_{0}ds\int^{\Lambda}_{0}dk_{1}
   \frac{k_{1}}{\omega_{\pi}(\vec{k_{1}})}\frac{2f_{\vec{k_{1}}}+1}{s} \nonumber \\
   &\times& \{ exp[i(x + \omega_{\pi}(\vec{k_{1}})- \omega_{s}(\vec{k}))s]
   + exp[-i(x + \omega_{\pi}(\vec{k_{1}}) + \omega_{s}(\vec{k}))s] \nonumber \\
   &+& exp[i(x - \omega_{\pi}(\vec{k_{1}}) - \omega_{s}(\vec{k}))s] 
   + exp[-i(x - \omega_{\pi}(\vec{k_{1}}) + \omega_{s}(\vec{k}))s] \}|_{x = \omega_{m}}^{\omega_{p}},
\end{eqnarray}
where
\begin{eqnarray}
      && f_{\vec{k_{1}}} = \frac{1}{e^{\omega_{\pi}(\vec{k_{1}})/T}-1},  \\
      && \omega_{\pi}(\vec{k}) = \sqrt{\vec{k}^2+m^{2}_{\pi}}, \\
      && \omega_{p} = \sqrt{(|\vec{k}| + |\vec{k_{1}}|)^2 + m_{\pi}^2},\\
      && \omega_{m} = \sqrt{(|\vec{k}| - |\vec{k_{1}}|)^2 + m_{\pi}^2},
\end{eqnarray}
and $\Lambda$ is the momentum cutoff.
The qualitative behavior of Im$A(\vec{k},t)$ 
is shown in FIG.1(a). 
One sees that Im$A(\vec{k},t)$ has an ultraviolet divergence for $t > 0$.
(Re$A(\vec{k},t)$ which corresponds to the decay width of 
$\sigma$ meson is finite.)

 In order to solve the coupled equations (\ref{eqn:sigma2pi}) 
and (\ref{eqn:sigma2pi2}), one has to remove the ultraviolet divergence 
by tuning the mass counter term. 
We choose $t_{0} = \infty$, that is, 
we require that the frequency shift 
of $\sigma$ meson (${\Delta}\omega$) vanishes 
at $\sigma$ meson momentum $\vec{k}=0$, $T = 0$ and $t = \infty$:
\begin{eqnarray}
   {\Delta}\omega = [{\rm Im}A(0,\infty) 
- \frac{1}{2}{\delta}m^{2}_{s0}(\infty)]/\omega(0)|_{T = 0} = 0. \label{eqn:fs}
\end{eqnarray}
By this procedure, the imaginary part of 
the renormalized $2\pi$-$2\pi$ correlation function defined by 
${\rm Im}A_{ren}(\vec{k},t) \equiv 
{\rm Im}A(\vec{k},t) - \frac{1}{2}\delta m^2_{s0}(\infty)$
becomes finite for $t > 0$, as shown in FIG.1(b).
The divergence of ${\rm Im}A(\vec{k},t)$ for $t>0$ is transfered to 
the divergence at $t=0$ after renormalization.
 In order to see the cutoff dependence, 
we have carried out numerical calculation by taking $m_{s} = 550MeV$, 
$m_{\pi} = 140MeV$, $g = 1650MeV$, $k = 30MeV$ and $ T = 0$, i.e., 
$f_{k_{1}} = 0$.
The result for ${\rm Im}A(\vec{k},t)$ is shown 
in FIG.2. 
At large $t$, ${\rm Im}A(\vec{k},t)$ approaches rapidly 
to $\Lambda$-independent limit 
while an appreciable $\Lambda$-dependence remains near $t = 0$ 
as it diverges at $t=0$ as $\Lambda \longrightarrow \infty$.

 One may still worry about the presence of divergence at $t = 0$.
However, we can show that it is harmless 
in calculating physical quantities.
The reason is that the important quantity in solving (\ref{eqn:sigma2pi}) is 
not $A(\vec{k},t)$ itself but the integral $\int^{t}_{0}dsA(\vec{k},s)$ 
which is finite.
In fact, one can show that ${\rm Im}A(\vec{k},t)$ behaves like ${\rm ln}{\alpha}t$ 
for small but non-zero $t$:
\begin{eqnarray}
   \lefteqn{ \int^{t}_{0}ds\int^{\Lambda}_{0}dk_{1}\frac{k_{1}}{s\omega_{\pi}(\vec{k_{1}})}
[ {\rm cos}(\omega_{p} + \omega_{\pi}(\vec{k_{1}}) - \omega_s(\vec{k}))s} \nonumber \\
&& - {\rm cos}(\omega_{m} + \omega_{\pi}(\vec{k_{1}})- \omega_s(\vec{k}))s] 
\sim \int^{t}_{0}ds\int^{\Lambda}_{0}dk_{1}\frac{-2}{s}
{\rm sin}(ks){\rm sin}(2k_{1}s) \nonumber \\
&& \hspace{5.3cm}\sim \int^{t}_{0}ds\frac{k}{s}({\rm cos}2{\Lambda}s - 1)  \nonumber \\
&& \hspace{5.3cm}\sim k \{\frac{{\rm sin}(2{\Lambda}t)}{2{\Lambda}t}
 - \frac{{\rm cos}(2{\Lambda}t)}{(2{\Lambda}t)^2}-{\rm ln}2{\Lambda}t -\gamma\}, \label{eqn:dcA3} 
\end{eqnarray}
where $\gamma$ is the Euler constant.
In going from the first line to the second line on the r.h.s.,
we use the condition that 
$k << 1/t$ and from the second line to the third line, $\Lambda t>> 1$.
In the limit of $\Lambda \longrightarrow \infty$, 
we have
\begin{eqnarray}
 {\rm the~ last~ line~ of~~} (\ref{eqn:dcA3})~ &\sim& -k{\rm ln}~2{\Lambda}t \nonumber \\
&=& -k({\rm ln}~\alpha^{-1}{\Lambda} + {\rm ln}~\alpha{t}),
\end{eqnarray}
where $\alpha$ is a constant which has a mass dimension .
The first term is an ultraviolet logarithmic divergence which can be renormalized by
the mass counter term.
The second term is a new divergence  
which cannot be eliminated by our procedure.
Fortunately, this new divergence is logarithmic.
So the integral $\int^{t}_{0}dsA(\vec{k},s)$ is finite,
and one gets a convergent result 
for $\langle c^{\dagger}_{\vec{k}}(t) \rangle$ in 
the limit $\Lambda \longrightarrow \infty$.

Furthermore it is remarkable that there is no new divergence when $T \neq 0$ 
if we renormalize it at $T = 0$ like in  usual equilibrium thermal field theory.
This can be seen in FIG.3.

Combining the above results, we can evaluate 
Re$\langle c^{\dagger}_{\vec{k}}(t) \rangle$ for various $T$,
of which result is shown in FIG.4. 
Starting from an arbitrarily chosen initial value, 
$\langle c^{\dagger}_{\vec{k}}(t) \rangle$ makes a damped oscillation towards 
a equilibrium value which is vanishing for $|\vec{k}| \neq 0$ in this order.

To summarize, we have calculated time dependent correlation functions 
using Shibata-Hasitsume's projection operator method.
We encounter a divergence which can be renormalized by a mass counter term. 
The remaining divergence at the initial time $t=0$ is harmless 
in calculating physical quantities.
It remains as a future problem to study the role 
of wave function renormalization and vertex renormalization
which will be necessary in more complicated cases or higher order calculations.

Anastopoulos \cite{ref:13} has recently studied a similar problem by using 
Nakajima-Zwanzig's projection operator method.
He uses a different projection operator 
and a different renormalization procedure.

\newpage

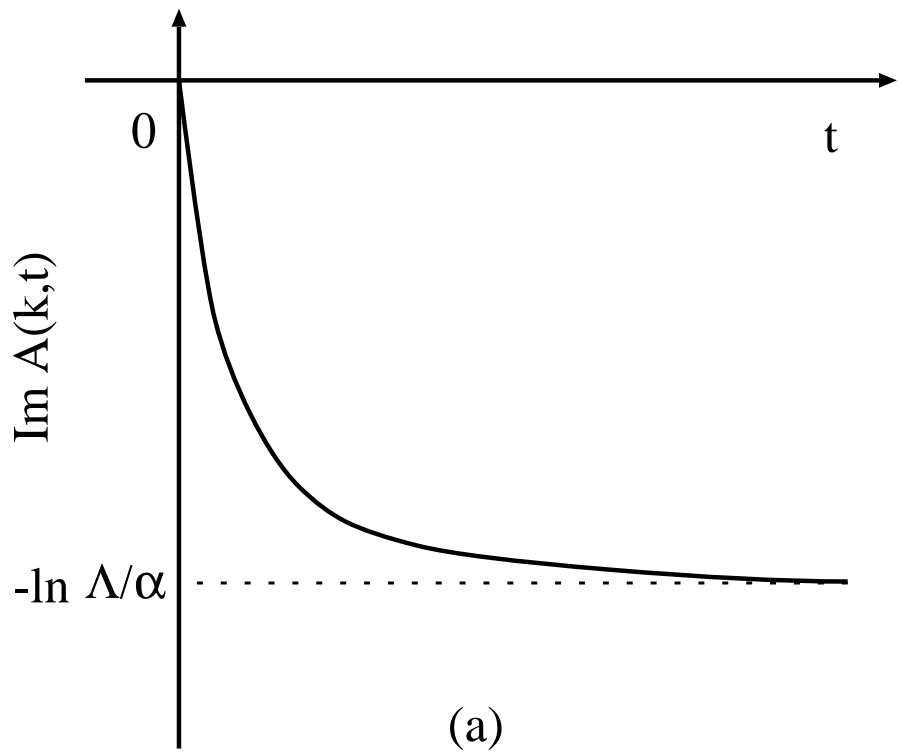
FIG.1:Qualitative behaviors of Im$A(\vec{k},t)$ before ((a)) and after ((b)) 
renormalization.
\\

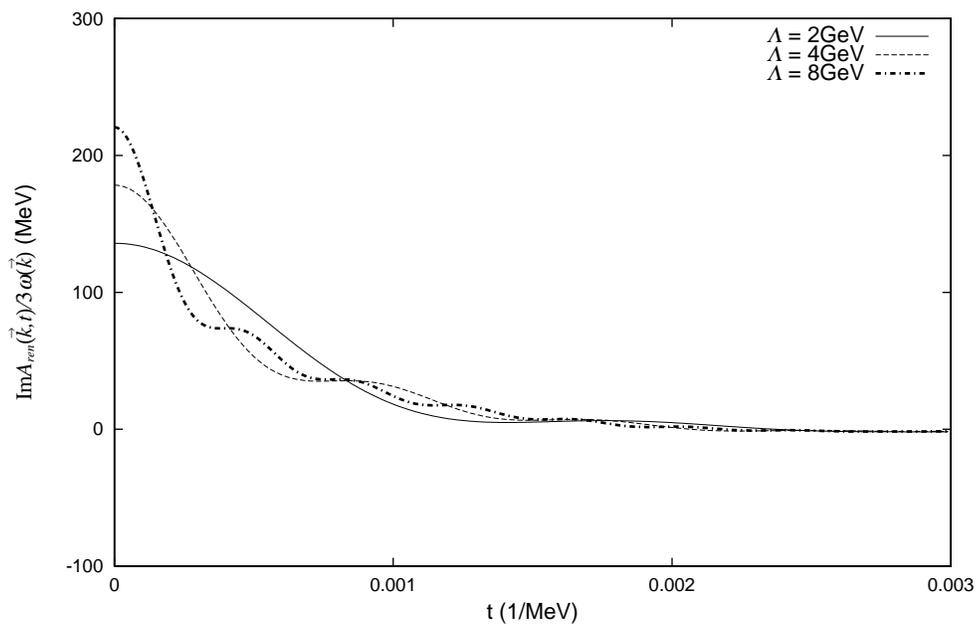
FIG.2:Cutoff dependence of Im$A_{ren}(\vec{k},t)$.
Solid, dashed and dot-dashed lines are the results 
for momentum cutoff $\Lambda = 2GeV$, $4GeV$ 
and $8GeV$, respectively, at $T = 0MeV$.
\\

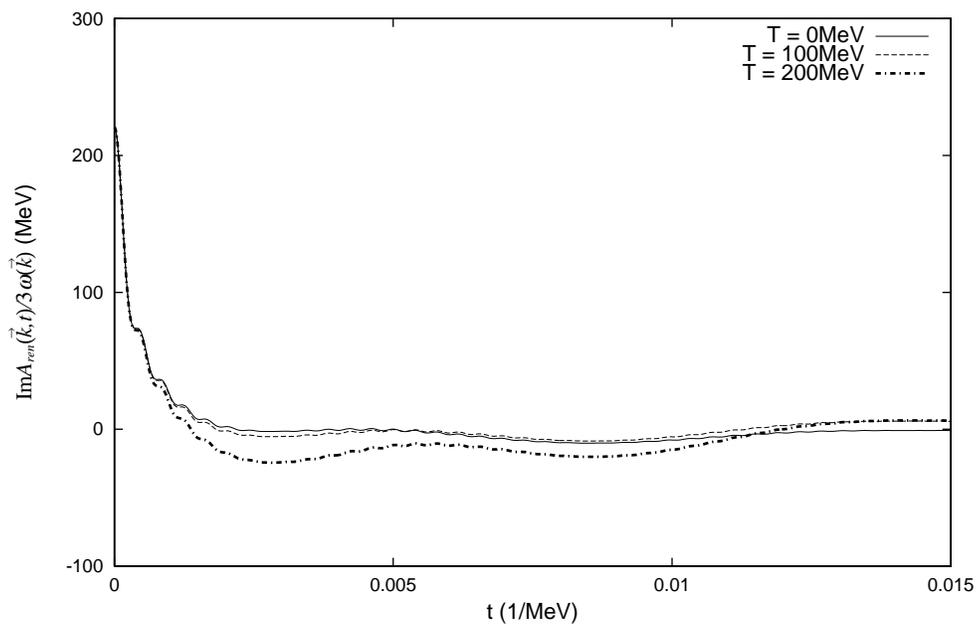
FIG.3:Temperature dependence of Im$A_{ren}(\vec{k},t)$ for $\Lambda = 8GeV$.
Solid, dashed and dot-dashed lines are the results for temperature $T = 0MeV$, $100MeV$ and $200MeV$, respectively.
\\

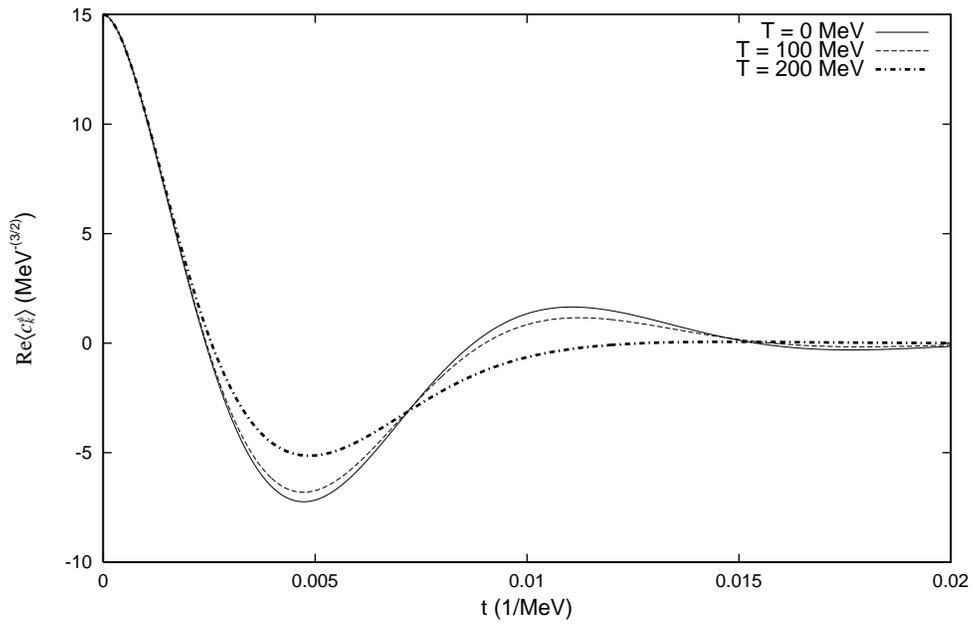
FIG.4:Temperature dependence of 
Re$\langle c^{\dagger}_{\vec{k}}(t) \rangle$.
Solid, dashed and dot-dashed lines are the results for temperature $T = 0MeV$,
$100MeV$ and $200MeV$, respectively, and 
$\langle c^{\dagger}_{\vec{k}}(0) \rangle = 15MeV^{-\frac{3}{2}}$.

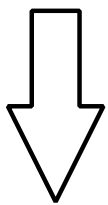
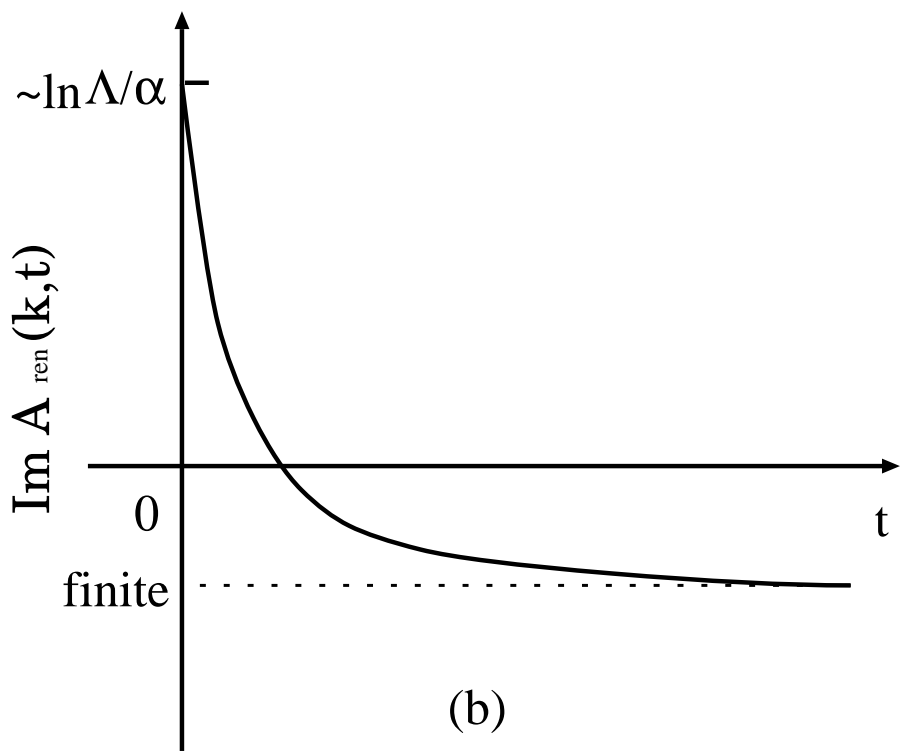
\end{document}